\documentclass[12pt,english]{article}

\usepackage[maccyr]{inputenc}

\usepackage{amssymb,amsmath}
\usepackage{cite}
\usepackage{color}
\begin{document}
\begin{titlepage}
\samepage{
\setcounter{page}{1}

\begin{center}
{\Large \bf \vspace{3cm} Asymmetric Orbifolds, Non-Geometric Fluxes and \\ \vspace{3mm} Non-Commutativity in Closed String Theory}

\vspace{17mm}
\begin{minipage}{0.97\linewidth}
\begin{center}
 {\normalsize\bf Cezar Condeescu$^{1}$,~ Ioannis Florakis$^{2,3}$ and Dieter L\"{u}st$^{2,3}$ \\
}
\vspace{10mm}

\end{center}
\vspace{.3cm} \hspace{2.0cm}
\begin{minipage}{1.25\linewidth}
\begin{footnotesize}
\begin{itemize}
	\item[$^{1}$]   Department of Theoretical Physics \\
 Horia Hulubei National Institute of Physics and Nuclear Engineering \\
 P.O. Box MG-6, M\u{a}gurele - Bucharest, 077125, Jud. Ilfov, Rom\^{a}nia    
	\item[$^{2}$]  Arnold Sommerfeld Center for Theoretical Physics\\
	 Fakult\"at f\"ur Physik, Ludwig-Maximilians-Universit\"at M\"unchen \\
	 Theresienstr. 37, 80333 M\"unchen, Germany  
	 \item[$^{3}$]  Max-Planck-Institut f\"ur Physik,\\ Werner-Heisenberg-Institut, 80805 M\"unchen, Germany   
\end{itemize}
\end{footnotesize}
  \vspace{1cm}
 \end{minipage}
\end{minipage}

\vspace{5mm}
\begin{abstract}\vspace{5mm}
{\normalsize  In this paper we consider a class of exactly solvable  closed string flux backgrounds that exhibit non-commutativity in the closed string coordinates. They are realized in terms of freely-acting asymmetric $\mathbb{Z}_N$-orbifolds, which are themselves close relatives of twisted torus fibrations with elliptic $\mathbb{Z}_N$-monodromy (elliptic T-folds). We explicitly construct the modular invariant partition function of the models and derive the non-commutative algebra  in the string coordinates, which is exact to all orders in $\alpha'$. Finally, we relate these asymmetric orbifold spaces to inherently stringy Scherk-Schwarz backgrounds and non-geometric fluxes.
}
\end{abstract}
\end{center}

\smallskip}

\vspace*{-21cm}
\begin{flushright}
LMU-ASC 12/12\\
MPP-2012-12
\end{flushright}

\vfill
{\small
\begin{itemize}
\item[E-mail:] {\tt ccezar@theory.nipne.ro}\\ {\tt florakis@mppmu.mpg.de}\\
{\tt dieter.luest@lmu.de}
\end{itemize}
}

\end{titlepage}

\vfill


\setlength{\topmargin}{-1.1cm}
\section{Introduction}
\bigskip

It has been known since the work of \cite{Chu:1998qz, Schomerus:1999ug, Seiberg:1999vs,Ardalan:1998ks} that non-commutative geometry can naturally arise in open string compactifications by turning on a constant $B$-field
or an abelian magnetic field on the worldvolume of a D-brane. In particular, the presence of non-trivial magnetic flux leads to a shift in the frequencies of
string oscillators and, hence, to non-trivial commutation relations for the open string coordinates. Furthermore, it was argued \cite{Seiberg:1999vs} that the
low energy effective theory of such compactifications admits an equivalent description in terms of  a gauge theory on a non-commutative space.

On the other hand, in the case of closed strings, turning on a constant $B$-field typically preserves the commutativity of the coordinates. However, it has been
pointed out in recent works \cite{Lust:2010iy,Blumenhagen:2010hj,Blumenhagen:2011ph,Blumenhagen:2011yv} that non-commutativity or even non-associativity may arise in closed string compactifications in the presence of non-trivial (non-geometric) fluxes. The starting point in these works was to consider special flux backgrounds, in particular, non-trivial  fibrations of a two-torus over a one-dimensional circle. Whenever the transition functions of this fibration are the standard diffeomorphisms, one is dealing with geometrical flux backgrounds. If, however, the transition
functions also include some stringy T-duality transformation, the corresponding backgrounds have only locally the structure of a Riemannian manifold with fluxes. From a global perspective, these are no longer manifolds but, rather, fall into the class of  so-called T-folds \cite{Hull:2004in} and one speaks about non-geometric flux backgrounds ($Q$-fluxes) \cite{Shelton:2005cf, Dabholkar:2005ve}. Moreover, there is reason to assume that there exist other, more exotic non-geometric string backgrounds ($R$-fluxes), that do not even admit a local Riemannian description.
As shown in \cite{Lust:2010iy,Blumenhagen:2010hj,Blumenhagen:2011ph,Blumenhagen:2011yv}, the closed string coordinates for these classes of non-geometric backgrounds typically become non-commutative or even non-associative.

On many occasions, it so happens that geometric and non-geometric  backgrounds can be related to each other by T-duality. In fact, a particular class of three-dimensional torus fibrations
exhibits a nice chain of three T-dualities \cite{Shelton:2005cf}, which connect
four different background spaces: the flat torus with constant $H$-flux (the $H$-background),
the twisted torus with geometric metric $\omega$-flux (the $\omega$-background), the non-geometric T-fold
with $Q$-flux and, finally, a more speculative background with $R$-flux:
 \begin{align}
  H_{abc}\rightarrow f^{a}{}_{bc}\rightarrow Q_{a}{}^{bc}\rightarrow R^{abc}\;.
 \end{align}
In  $Q$-flux backgrounds the coordinates of the fiber torus do not commute, while in the case of the $R$-flux background the full non-associative structure among all three
coordinates becomes visible.  This structure was nicely illustrated in \cite{Lust:2010iy}, where a three-torus with so-called elliptic $\mathbb{Z}_4$-monodromy was investigated: the coordinates
of the geometric $\omega$-background are clearly commuting, whereas the T-dual torus coordinates of the non-geometric $Q$-backgrounds are non-commuting, with
the dual momentum of the circle direction appearing on the right hand side of the commutation relation. Note that, in the first case, the $\mathbb{Z}_4$-monodromy acts left-right
symmetrically on the coordinates of the two-torus, whereas in the latter, T-dual case, the $\mathbb{Z}_4$-action on the torus is left-right asymmetric.
Hence, one recognizes that closed string non-commutativity  and non-associativity generically depend on the chosen T-duality frame and is not a T-duality invariant notion per se. This suggests that the framework of doubled field theory \cite{Hull:2009mi}, where one is considering coordinates as well as dual coordinates on
more or less equal footing, is a suitable framework for studying non-commutativity  and non-associativity in closed string theory.\footnote{Recently, a geometric ten-dimensional action for
non-geometric $Q$- and $R$-flux backgrounds was derived using doubled field theory resp. by considering field redefinitions that are motivated by T-duality
\cite{allp11,new}.}

From the closed string perspective, the emergent non-commutative behavior in the coordinates is, certainly, not a surprise.
Indeed, string theory at curvatures of the order of the string scale implies a fundamentally different picture for spacetime than the one expected from the field theory approximation. It is well-known (see, for example, \cite{Kiritsis:1994mu,Kiritsis:1995wb} and references therein) that classical notions such as geometry, topology and even space dimensionality itself only become unambiguously defined in the effective low-energy approximation. In the strong curvature regions, on the other hand, field-theoretic notions typically break down and new inherently stringy phenomena occur. This purely stringy behavior is essentially related to the presence of extended symmetry points in the moduli space of the theory, where the contribution of winding states conspire together with momentum modes in order to form additional massless states. As a result, a proper description of these regimes cannot rely on the effective supergravity approximation and requires, instead, a description in terms of an exact conformal field theory (CFT).

So far, all the above-mentioned approaches share the limitation that the (non)-geometric backgrounds considered are  solutions to the string equations of motion only to the lowest-order in $\alpha'$ and, in general, do not correspond to solvable string backgrounds\footnote{A notable exception is \cite{Blumenhagen:2010hj}, where the authors consider the solvable $SU(2)_k$ WZW background, which is indeed exact to all orders in $\alpha'$. However, their result for the non-associative algebra is effectively derived only as a lowest-order expansion in the curvature and, hence, in $\alpha'$.}. The purpose of this paper is to provide the first examples of solvable flux backgrounds, which are consistent and exact to all orders of $\alpha'$, and which exhibit an exactly calculable non-commutative structure in the closed string coordinates.

The starting point is a non-geometric analogue of the twisted torus (see also \cite{Lust:2010iy}) with elliptic $\mathbb{Z}_N$-monodromy, corresponding to a $Q$-flux background. At special points of the moduli space, this non-geometric background  can be described by a freely-acting asymmetric $\mathbb{Z}_N$-orbifold \cite{Narain:1986qm}. Since the models are now fully solvable, we can prove that, in the case of asymmetric orbifolds, the coordinates of the two-torus are indeed non-commuting. The non-commutative structure constants of the algebra are neatly parametrized by a Scherk-Schwarz \cite{Scherk:1978ta} flux-matrix $F_{IJ}$ encoding the non-trivial monodromy properties, and are crucially dependent on the winding number in the circle direction. Therefore, our result now fully establishes the presence of non-commutativity in closed string flux compactifications. We further reinterpret these non-geometric backgrounds  as a new class of stringy `asymmetric' Scherk-Schwarz compactifications with non-geometric flux. Moreover, it is important to note that the asymmetric orbifolds we construct are not T-dual to symmetric ones and, in this sense, they can be considered as `truly' asymmetric\footnote{The importance of such constructions was noted, for example, in \cite{Wecht:2007wu}. For further discussion on the relation between asymmetric orbifolds and non-geometric backgrounds see also \cite{Flournoy:2004vn}.}.

The paper is organized as follows. In Section \ref{SectionMotivation} we recall some facts about twisted torus compactifications, the r\^{o}le of monodromies and T-duality and
how closed string non-commutativity appears in this class of background spaces.
In Section \ref{SectionOrbifolds} we introduce freely-acting (a)symmetric $\mathbb{Z}_N$-orbifolds as limiting solvable cases of the more general twisted tori and construct their modular
invariant partition functions. In Section \ref{SectionNonCommut}, we proceed to obtain the non-commutative coordinate algebra for this class of freely-acting asymmetric orbifolds. Finally, in Section \ref{ScherkSchwarzSection}, we relate
the above backgrounds to `asymmetric' Scherk-Schwarz compactifications with  non-geometric fluxes.


\section{Torus fibrations, T-folds and Non-commutativity}\label{SectionMotivation}
\bigskip

In this section we review some basic notions of generalized Scherk-Schwarz compactifications and their relation to T-folds and closed string non-commutativity.
In certain cases, it is known that such backgrounds may admit an exact CFT description in terms of freely-acting orbifolds and, hence,
they may provide useful tools for studying string theory in such non-geometric setups\footnote{Our convention throughout the text will be $\alpha'=1$.}.

String theory compactified on a $N$-dimensional torus $T^N$ has a T-duality symmetry given by the group $O(N,N;\mathbb{Z})$, of  $2N$-dimensional matrices $g$ with
 integer entries preserving the indefinite $O(N,N;\mathbb{Z})$-invariant metric $L$:
\begin{align}
g^T Lg=L \qquad {\rm  with\ }\qquad L=\left(
                                     \begin{array}{cc}
                                       0 & 1_N \\
                                       1_N & 0 \\
                                     \end{array}
                                   \right)~.
\end{align}
The matrices $g$ can be parametrized by the $d$-dimensional matrices $a,b,c,d$ :
\begin{align}
g=\left(
    \begin{array}{cc}
      a & b \\
      c & d \\
    \end{array}
  \right)~,
  \label{g-twist}
\end{align}
satisfying the following conditions:
\begin{align}
\begin{split}
	&a^T c+c^T a=0  ~,\\
	&a^T d+c^T b=1_N ~,\\
	&b^T d+d^T b=0 ~.
\end{split}
\label{conditions}
\end{align}

The action of  T-duality on the metric $G$, antisymmetric tensor $B$ and dilaton $\Phi$ is non-linear and is given by the Buscher rules \cite{Buscher:1987sk}:
\begin{align}
\begin{split}
	&E' = (a E+b)(c E+d)^{-1}~,\\
	& \sqrt{\det{G'}} ~e^{-2\Phi'} = \sqrt{\det{G}} ~e^{-2\Phi}~,\\
\end{split}\label{Buscher}
\end{align}
with $E=G+B$. These were derived by using methods of gauged sigma models \cite{Buscher:1987sk,Rocek:1991ps,Giveon:1991jj,Giveon:1994fu}, in which different
gauges give rise to different T-dual backgrounds. Strictly speaking, the existence of a global isometry which can be gauged is crucial for this procedure to be carried out.
However, it has been argued \cite{Hull:2006qs} that, in certain cases, one might still be allowed to perform a T-duality, using the Buscher rules \eqref{Buscher}, even in the
absence of such isometries.

In a (generalized) Scherk-Schwarz mechanism one compactifies further on a circle $S^1$ and the reduction can then be twisted by a non-trivial $O(N,N;\mathbb{Z})$ monodromy, corresponding to
the T-duality symmetry of string theory on $T^N$. In this case, the fields are allowed to carry non-trivial dependence on the circle coordinate, $\mathbb{X}$. The most common example is that of the twisted torus. It corresponds to a $T^N$-torus fibered over $S^1$ with monodromy of the form:
\begin{align}
g_a=\left(
      \begin{array}{cc}
        a & 0 \\
        0 & (a^T)^{-1} \\
      \end{array}
    \right)~.
\end{align}
This monodromy matrix defines the embedding of the geometric subgroup $GL(N;\mathbb{Z})$ into the T-duality group $O(N,N;\mathbb{Z})$. Notice that an element
$a\in GL(N;\mathbb{Z})$ corresponds to a large diffeomorphism of the $T^N$-fiber. Indeed, going once around the $S^1$-circle, i.e. under the transformation
$\mathbb{X}\rightarrow \mathbb{X}+2\pi$, we see that $E$ transforms as:
\begin{align}
 E'=g_a E g_a^T~,
\end{align}
which corresponds to a transition between two patches in the base space $S^1$. Hence, the total space of the bundle, for twists inside the
subgroup $GL(N;\mathbb{Z})$, defines a classical geometric string background and one can still distinguish between the metric $G$ and $B$-field, with the corresponding tranformations being given by $G'=g_a G g_a^T$ and $B'=g_a B g_a^T$.

The example of a twisted three-torus $T^3$ has been considered often in the literature as a toy model to illustrate how non-commutative geometry \cite{Lust:2010iy}
can arise in string theory or to argue about the existence of more general non-geometric fluxes \cite{Shelton:2005cf,Dabholkar:2005ve}(e.g. $Q$-fluxes and $R$-fluxes). Let us consider a torus $T^2$ fibered over $S^1$ with coordinates $X^1, X^2$ on the fiber and $\mathbb{X}$ on the base\footnote{Strictly speaking, a twisted three torus is not in itself a good background for string theory. One should, rather,  fiber this construction over some appropriate manifold in order to obtain a consistent background.}.  The metric of the total space is of the of the following form:
\begin{align}
ds^2=\frac{1}{ \tau_2(\mathbb{X})}\left|dX^1+\tau(\mathbb{X})dX^2\right|^2+d\mathbb{X}^2~,
\end{align}
with the complex structure $\tau=\tau_1+i\tau_2$ of the fiber being, in general, a non-trivial function of the base space coordinate.\footnote{When going to the (a)symmetric orbifold point in the next chapter, we will set $\tau=i$, i.e.
the background becomes flat.}

Performing a T-duality in the $X^1$ direction takes one to a string theory compactified on a $T^3$-torus with non-trivial $H$-flux. Indeed, making use of the Buscher rules one obtains a $B$-field of the form:
\begin{align}
B=\tau_1(\mathbb{X})~ dX^1\wedge dX^2~,
\end{align}
where the function $\tau(\mathbb{X})$ has now the interpretation of a K\"ahler modulus and the corresponding $H$-flux is given by:
\begin{align}
H=\partial_{\mathbb{X}}\,\tau_1(\mathbb{X})~ dX^1\wedge dX^2\wedge d\mathbb{X}~.
\end{align}

Let us now consider a simple example where the T-dual theory contains  constant $H$-flux. This can arise from a parabolic monodromy $g_a$, with $a\in GL(2;\mathbb{Z})$ corresponding to integer translations:
\begin{align}
a=\left(
    \begin{array}{cc}
      1 & m \\
      0 & 1 \\
    \end{array}
  \right) \quad,\quad m\in\mathbb{Z}~.
\end{align}
Indeed, the T-dual $B$-field and corresponding $H$-flux will now take the form:
\begin{align}
B=\frac{m\mathbb{X}}{2 \pi}dX^1\wedge dX^2\quad, \quad H=\frac{m}{2\pi}dX^1\wedge dX^2\wedge d\mathbb{X}~.
\end{align}

Another important class of torus fibrations is given by elliptic monodromies, which act on the toroidal coordinates as rotations. For example, consider the monodromy corresponding to a $\mathbb{Z}_4\subset O(2;\mathbb{Z})$ rotation:
\begin{align}\label{z4}
\left(
  \begin{array}{c}
    X^1 \\
    X^2 \\
  \end{array}
\right) \rightarrow \left(
  \begin{array}{c}
    X^2 \\
    -X^1 \\
  \end{array}
\right)~,
\end{align}
resulting in an $SL(2;\mathbb{Z})$-transformation of the complex structure of the fiber:
\begin{align}
\tau(\mathbb{X})\rightarrow -1/\tau(\mathbb{X})~.
\end{align}
Notice that, for $\tau(\mathbb{X})=i$, the complex structure is a fixed point of the above transformation. At this point in moduli space, the fibered torus admits an exact CFT description in terms of a freely-acting $\mathbb{Z}_4$-orbifold, corresponding to the minimum of the Scherk-Schwarz potential for the complex structure, \cite{Dabholkar:2002sy, Hull:2005hk}.

One can diagonalize the $\mathbb{Z}_4$ rotation (\ref{z4}) by introducing complex coordinates $Z=\frac{1}{\sqrt{2}}(X^1+iX^2)$ and, hence, obtain the following twisted boundary conditions:
\begin{align}\label{IntroBC}
Z(\tau,\sigma+2\pi)=e^{2\pi i \theta}Z(\tau,\sigma)~,
\end{align}
with the angle $\theta$ depending on the winding number $n^{\mathbb{X}}$ in the $S^1$-direction:
\begin{align}
\theta=-fn^{\mathbb{X}} \quad, \quad {\rm with\ }~~ f\in\tfrac{1}{4}+\mathbb{Z}~.
\end{align}
Even though the complex structure $\tau(\mathbb{X})$ (or the $H$-field, in the T-dual picture) is a non-trivial function of the base coordinate $\mathbb{X}$ and the $\sigma$-model is, in general, non-linear, one could still  write down a mode expansion for the fiber coordinates subject to the twisted boundary conditions (\ref{IntroBC}), with the understanding that the result is only a lowest-order approximation in the $\omega$- or $H$-flux (and in $\alpha'$). Introducing the usual left- and right-moving coordinates $Z_{L,R}$, one obtains:
\begin{align}
\begin{split}
&Z_L(\tau+\sigma)=\frac{i}{\sqrt{2}}\sum_{k\in\mathbb{Z}}\frac{\alpha_{k-\theta}}{k-\theta}e^{-i(k-\theta)(\tau+\sigma)}~,\\
&Z_R(\tau-\sigma)=\frac{i}{\sqrt{2}}\sum_{k\in\mathbb{Z}}\frac{\tilde\alpha_{k+\theta}}{k+\theta}e^{-i(k+\theta)(\tau-\sigma)}~.
\end{split}
\end{align}
Similar expansions hold for the complex conjugates, $\bar Z_{L,R}$. The usual quantization procedure then leads to the familiar bosonic oscillator algebra for the complex-conjugate Fourier modes:
\begin{align}
[\alpha_{k-\theta},\bar\alpha_{\ell-\theta}]=(k-\theta)\delta_{k,\ell}~.
\end{align}
Explicit calculation then yields the equal-$\tau$ commutation relations for the coordinates\footnote{Strictly speaking, the commutation algebra should include the metric factor $G^{z\bar z}$, but this is immaterial for our present discussion and we will simply suppress it.}:
\begin{align}\label{IntroSeries}
[Z_L(\tau,\sigma),\bar Z_L(\tau,\sigma')]=\frac{1}{2}\sum_{k\in\mathbb{Z}}\frac{e^{-i(k-\theta)(\sigma-\sigma')}}{k-\theta}\equiv \frac{1}{2}\Theta(\sigma-\sigma',\theta)~.
\end{align}
For the right-moving coordinates, a similar calculation yields  the same function $\Theta$, but with the opposite sign:
\begin{align}
[Z_R(\tau,\sigma),\bar Z_R(\tau,\sigma')]=-\frac{1}{2}\Theta(\sigma-\sigma',\theta)~.
\end{align}
In order to obtain a local result, one should carefully investigate the limit $\sigma\rightarrow \sigma'$. Since in this limit, the series representation (\ref{IntroSeries}) is naively divergent, it has to be defined through analytic continuation.
To this end, upon introducing the complex variable $z=e^{i(\sigma'-\sigma)}$, the function $\Theta$ can be neatly represented in terms of hypergeometric functions:
\begin{align}\label{hypergeometric}
\Theta(\sigma-\sigma',\theta)=-\frac{z^{-\theta}}{\theta}\Big[ {}_2F_1(1,-\theta;1-\theta;z)+ {}_2F_1(1,\theta;1+\theta;z^{-1})-1\Big]~,
\end{align}
which can then be analytically continued to $z\rightarrow 1$. The final result for $\Theta$ is given in terms of elementary functions:
\begin{align}
\Theta(\theta) \equiv \left\{ \begin{array}{c l}
			-\pi\cot (\pi\theta) & , \quad \theta\notin \mathbb{Z} \\
			0 & , \quad \theta \in \mathbb{Z} \\
				\end{array}\right. ~,
\end{align}
modulo the discontinuity at $\theta\in\mathbb{Z}$, arising from the subtraction of the zero mode in (\ref{IntroSeries}).

As expected for a geometric background, the twisted torus leads to commutative coordinates:
\begin{align}
[X^1(\tau,\sigma),X^2(\tau,\sigma)]=0~,
\end{align}
where we have expressed the commutation relations in terms of the original coordinates $X^1,X^2$ of the torus fiber.
Nevertheless,  performing a T-duality in the $X^1$-direction, one is lead to a non-commutative coordinate algebra. Indeed, introducing the dual coordinate $\tilde X^1=X^1_L-X_R^1$, one obtains:
\begin{align}
[\tilde X^1(\tau,\sigma),X^2(\tau,\sigma)]=[X_L^1(\tau,\sigma),X_L^2(\tau,\sigma)]-[X_R^1(\tau,\sigma),X_R^2(\tau,\sigma)]=i\,\Theta(\theta)~.
\end{align}

The same commutation algebra arises by imposing from the very beginning the following asymmetric boundary conditions:
\begin{align}
\begin{split}
&Z_L(\tau,\sigma+2\pi)=e^{2\pi i\theta}Z_L(\tau,\sigma)~,\\
&Z_R(\tau,\sigma+2\pi)=e^{-2\pi i\theta}Z_R(\tau,\sigma)~,\\
\end{split}
\label{asymmetric}
\end{align}
which are highly reminiscent of an asymmetric orbifold\footnote{More general asymmetric rotations are possible for other choices of $\theta_L$ and $\theta_R$, provided they are compatible with modular invariance. In this case, the algebra of coordinates becomes $[X^1,X^2]=\frac{i}{2}\left\{\Theta(\theta_L)-\Theta(\theta_R)\right\}$.}. The first line in eq. (\ref{asymmetric}) corresponds to the $\mathbb{Z}_4$-action:
\begin{align}
X_L^1\rightarrow X_L^2 \quad, \quad X_L^2\rightarrow -X_L^1~,
\end{align}
whereas the second line now gives for the right-moving sector:
\begin{align}
X_R^1\rightarrow -X_R^2 \quad, \quad X_R^2\rightarrow X_R^1~.
\end{align}
These asymmetric rotations define the following monodromy element $g$ of the $O(2,2;\mathbb{Z})$ T-duality group:
\begin{align}
g =\left(
            \begin{array}{cccc}
              0 & 0 & 0 & 1 \\
              0 & 0 & -1 & 0 \\
              0 & 1 & 0 & 0 \\
              -1 & 0 & 0 & 0 \\
            \end{array}
          \right)~.
\end{align}
Notice that $g$ is no longer an element of the geometric subgroup $GL(2;\mathbb{Z})$. The arguments presented above, hence, support the following conclusion. In stringy Scherk-Schwarz compactifications, non-commutativity for the closed string coordinates can arise whenever the monodromy matrix $g\in O(N,N;\mathbb{Z})$ is not an element of the geometric subgroup $GL(N;\mathbb{Z})$. Furthermore, for special points in the moduli space of the theory, an exact CFT description in terms of freely-acting asymmetric orbifolds may exist for such non-geometric models. Of course, the construction of asymmetric orbifolds is not automatic. In general, modular invariance severely constrains the space of consistent asymmetric orbifold vacua (see for example \cite{Aoki:2004sm}). In the next sections we will present explicit constructions of freely-acting asymmetric toroidal orbifolds and show  how non-commutativity arises in these setups, by determining the structure constants of the non-commutative algebra and relating them to non-geometric Scherk-Schwarz fluxes. In particular, our results will be exact to all orders in $\alpha'$.

Before ending this section, however, it will be instructive to comment on the relation of these constructions with the notion of T-folds. A T-fold\footnote{Other constructions are possible. We mention here the approach double field theory \cite{Hull:2009mi} which takes T-duality as a fundamental symmetry of the theory and the one of generalized geometry \cite{Hitchin:2000jd, Hitchin:2001rw, Hitchin:2004ut, Gualtieri:2004rw} which attempts to generalize classical geometry by including the B-field as a gerbe connection on a formal sum of the tangent and cotangent bundles.} \cite{Hull:2004in} provides a natural geometric interpretation for stringy Scherk-Schwarz compactifications with T-duality twists.
By definition, on a T-fold, the fields (e.g. metric $G$ and $B$-field) are defined globally only up to T-duality transformations. Given the monodromy matrix $g$ in (\ref{g-twist}), going once around the base space $S^1$ of the Scherk-Schwarz fibration implies:
\begin{align}
E(\mathbb{X}+2\pi)=(aE(\mathbb{X})+b)(cE(\mathbb{X})+d)^{-1}~.
\end{align}
One can then build a geometrical bundle by enlarging the fiber space to include, aside from the torus coordinates $X=(X^1,...,X^N)$, also the corresponding duals
$\tilde X=(\tilde X^1,..., \tilde X^N)$, such that the action of the monodromy matrix $g\in O(N,N;\mathbb{Z})$ on the doubled $T^{2N}$-torus is purely geometric:
\begin{align}
\left(
                  \begin{array}{c}
                    X' \\
                    \tilde X' \\
                  \end{array}
                \right) = \left(
  \begin{array}{cc}
    a & b \\
    c & d \\
  \end{array}
\right) \left(
          \begin{array}{c}
            X \\
            \tilde X \\
          \end{array}
        \right)~.
\end{align}
In principle, however, this treatment implies a doubling of the number of degrees of freedom. One should then introduce additional constraints on the $T^{2N}$-bundle which, essentially, amount to imposing that $X_{L,R}^I=X^I\pm\tilde X^I$ are indeed left- and right-moving coordinates. A consistent choice of $T^N$-subbundle will then define locally a notion of space-time. The interpretation of the double torus bundle is that the enlarged space includes now all possible T-duals and a different choice of duality frame (polarization) takes one from a given model to its T-dual.

The connection of T-folds to freely-acting orbifolds can also be seen in the following way. A symmetric orbifold is given by an (abelian) finite group $\Gamma_{sym}$, which is a subgroup of $O(N;\mathbb{Z})$ inside the geometric $GL(N;\mathbb{Z})$-group. In our discussion of twisted tori we have seen that the boundary conditions for  elliptic monodromies are indeed similar to those of a symmetric orbifold. On the other hand, in the case of asymmetric orbifolds one is dealing with a finite group $\Gamma_{asym}$, which is, instead, a subgroup of $O(N;\mathbb{Z})_L\times O(N;\mathbb{Z})_R$, which, in general, is not contained in the geometric $GL(N;\mathbb{Z})$-group. Hence, freely-acting asymmetric orbifolds yield non-trivial T-folds.




\section{Freely-acting Asymmetric Orbifolds}\label{SectionOrbifolds}
\bigskip

As outlined in the previous sections, the consistent study of non-commutative effects in string theory necessitates a treatment that is exact to all orders in $\alpha'$. Following the logic of the Section 
\ref{SectionMotivation}, we are lead to consider the propagation of strings in flux backgrounds which admit an exact CFT description, hence, allowing us to consistently study the emerging non-commutativity.

A natural candidate is provided by a class of freely-acting asymmetric $\mathbb{Z}_N$-orbifolds. The advantage of considering such backgrounds is that the corresponding worldsheet CFT is locally free, allowing us to obtain exact mode expansions for the internal coordinates and their commutators, $[X^I,X^J]$. As we will see in Section \ref{ScherkSchwarzSection}, due to their freely-acting structure, the orbifolds we construct admit a target-space interpretation in terms of non-geometric $U(1)$-flux  along the non-commutative directions. This is the generalization of the Scherk-Schwarz mechanism \cite{Scherk:1978ta} to string theory, \cite{Ferrara:1987qp,Kounnas:1989dk,Antoniadis:1998ep} (see also \cite{Kiritsis:1997ca}).

We will start with the construction of a simple class of freely-acting asymmetric orbifold models. To illustrate the consistency of the models, we will explicitly display their modular invariant partition function. Even though in the present section we will focus on the orbifold picture, the discussion of Section \ref{SectionMotivation} 
clearly implies that these models should be considered as special solvable points in the moduli space of more general, non-geometric $Q$-flux backgrounds. Hence, our results will be exact to all orders in $\alpha'$ and to all orders in the (quantized) value of the flux\footnote{Contrary to the case of non-freely acting orbifolds, in which the internal space is singular at fixed points, the orbifolds we construct are free of such singularities, due to their freely-acting nature.}.


\subsection{Construction of asymmetric $\mathbb{Z}_N$-orbifolds}\label{ZNOrbifoldConstruction}
\bigskip

The constructions we present will be Type II (freely-acting) asymmetric orbifold models with $4\leq\mathcal{N}_4< 8$ spacetime supersymmetry, compactified on $(S^1\times T^5)/\mathbb{Z}_N$. The action of the $\mathbb{Z}_N$ on $T^5$ will be specified below. Let us also note that the restriction to Type II theories is only a convenient choice, as our results can be extended to Heterotic theories in a straightforward fashion.

We will start our discussion by considering an $N$-dimensional torus $T^N\subseteq T^5$, that is locally factorized from an $S^1$-circle of radius $R$. We will then proceed by defining the action of $\mathbb{Z}_N$ on this manifold and derive various consistency conditions, such that the theory is a well-defined asymmetric orbifold. In fact, in what concerns the $T^N$, we will restrict our attention to asymmetric orbifolds where the $\mathbb{Z}_N$ acts non-trivially only on the left-moving degrees of freedom. To this end, the discussion in this section refers to the left-moving worldsheet fields. It is then easy to extend the construction to the right-moving sector in order to define symmetric orbifolds as well.

Another point concerns the dimensionality of the torus, which we keep arbitrary at this point. As we will discuss later in this section, asymmetric orbifolds are highly constrained by modular invariance and by the requirement that the left- and right-moving CFTs factorize. The possible constructions, hence, depend highly on the dimension $N$ of the torus.

Let us take the $N$-dimensional torus $T^N$ to be parametrized by coordinates $X^I$, with $I=1,\ldots, N\leq 5$ and, further denote the coordinate associated to the $S^1$-circle by $\mathbb{X}$. We will restrict ourselves to the case where the orbifold acts as a permutation $P$ (including a possible `reflection') only on the left-moving coordinates of $T^N$, while leaving the right-movers invariant. Furthermore, in order to eliminate fixed points and generate a free action, we will couple this to a shift both in the momenta and windings of the $S^1$-direction in the orbifold basis. The orbifold element can be expressed, hence, as:
\begin{align}\label{GeneralOrbifold}
	g = e^{2\pi iQ_L}\,\delta~,
\end{align}
where $Q_L$ is the generator of permutations (with possible reflections) of the left-moving $T^N$ coordinates and  $\delta$ is an order $1/N$-shift in the $S^1$-direction. Of course, since we are considering an asymmetric orbifold, the construction can only take place at special points in the moduli space of the theory, where the CFT factorizes and purely left-moving lattice isometries exist. For simplicity, the models will be constructed at the fermionic point, where chiral bosons can be defined through fermionization \cite{Antoniadis:1985az}.

We shall now define the orbifold action on the left-moving $X^I$-coordinates of $T^N$. Consider the permutations (with a possible reflection)  defined by the matrix:
\begin{align}\label{OrbifoldAction}
	P_{IJ}(\epsilon)=\left( \begin{array}{r r r r r r}
							0 & 1 & 0 & \ldots & 0 & 0 \\
							0 & 0 & 1 & \ldots & 0 & 0 \\
							\vdots & \vdots & \vdots & \ldots & \vdots & \vdots \\
							0 & 0 & 0 & \ldots & 0 & 1 \\
							\epsilon & 0 & 0 & \ldots & 0 & 0 \\
						\end{array}
				\right)_{IJ}=\delta_{I,J-1}+\epsilon\,\delta_{I,J+N-1}~~,~~\textrm{where}~~\epsilon=\pm 1~,
\end{align}
where we display the orbifold action on the relevant coordinates only. Notice that in the absence of reflection ($\epsilon=+1$), this corresponds to a $\mathbb{Z}_N$-orbifold. On the other hand, including a non-trivial reflection ($\epsilon=-1$), leads to an enhanced $\mathbb{Z}_{2N}$.

We can now pick a basis that diagonalizes $P(\epsilon)$ as follows. If $\{\lambda_I\}$ is the set of eigenvalues of $P(\epsilon)$, then we can define the linear combinations:
\begin{align}
	Z^I =  \frac{1}{\sqrt{N}}\sum\limits_{J=1}^{N}{(\lambda_I)^{J-1}\,X^{J}}~~,~~\textrm{with}~~I=1,\ldots,N~,
\end{align}
which are the eigenvectors of $P(\epsilon)$ with eigenvalues:
\begin{align}
	\lambda_I(\epsilon) = e^{2\pi i (I-1+\nu)/N}~~,~~\textrm{with}~~\nu(\epsilon)=\left\{
								\begin{array}{c r}
										0 & ~,~\textrm{for}~ \epsilon=+1 \\
										1/2 & ~,~\textrm{for}~ \epsilon=-1 \\
								\end{array}
	\right. ~.
\end{align}
Notice that the new coordinates $Z^I$ can be either real or complex depending on the corresponding eigenvalue $\lambda_I$.

We now fermionize the $T^N$ coordinates of the original basis as:
\begin{align}
	i\partial X^I = iy^I\omega^I~,
\end{align}
where $y^I,\omega^I$ are real (auxiliary) free fermions. The change of basis is then defined by the unitary matrix:
\begin{align}
 	U_{IJ}(\epsilon)= \frac{1}{\sqrt{N}}\left( \begin{array}{r r r r r r}
							1 & \lambda_1 & \lambda_1^2 & \ldots & \lambda_1^{N-2} & \lambda_1^{N-1} \\
							1 & \lambda_2 & \lambda_2^2 & \ldots & \lambda_2^{N-2} & \lambda_2^{N-1} \\
							\vdots & \vdots & \vdots & \ldots & \vdots & \vdots \\
							1 & \lambda_N & \lambda_N^2 & \ldots & \lambda_N^{N-2} & \lambda_{N}^{N-1} \\
						\end{array} \right)_{IJ} = \frac{1}{\sqrt{N}}\, e^{2\pi i(I-1+\nu)(J-1)/N}~.
\end{align}
Note that the $\epsilon$-dependence of this matrix arises only implicitly, through the appearance of $\nu=\frac{1}{4}(1-\epsilon)$ in the eigenvalues. Explicitly, the change of basis on the coordinates and their fermionic (worldsheet) superpartners is:
\begin{align}\nonumber
	& Z^I = {U^I}_J(\epsilon)\,X^J~,\\
	& \Psi^I = {U^I}_J(\epsilon)\,\psi^J~.
\end{align}
In this basis, the orbifold acts simply as a rotation by a phase:
\begin{align}\nonumber
	& Z^I_L \rightarrow \lambda_I(\epsilon) Z^I_L~,\\ \label{ZNbc}
	& \Psi^I_L \rightarrow \lambda_I(\epsilon) \Psi^I_L~.
\end{align}
The action of the orbifold on the auxiliary free fermions is:
\begin{align}\nonumber
	& y^I \rightarrow {P^I}_J(\epsilon)\,y^J~,\\
	& \omega^I \rightarrow {P^I}_J(1)\,\omega^J~.
\end{align}
Notice that the $\omega$-fermions always transform under the orbifold as a pure permutation ($\epsilon=1$), in order for their product $y^I\omega^I$ to correctly represent the orbifold action on the bosons $\partial X^I$. Hence, the correct basis redefinition for the auxiliary free fermions is:
\begin{align}\nonumber
	& Y^I = {U^I}_J(\epsilon)\,y^I~,\\
	& W^I = {U^I}_J(1)\, \omega^J~.
\end{align}
It is, again, crucial that the $\omega$-fermions change basis as if there was no reflection ($\epsilon=+1$ or $\nu=0$).

In the new basis $\{Z^I,\Psi^I\}$, the coordinates and their fermionic superpartners generically arise as complex worldsheet fields. Complex eigenvalues always come in pairs ($\lambda^{*}=\epsilon\lambda^{N-1}$), so that we can group the associated coordinates such that $Z^{I},Z^{J}$ are complex conjugate to each other and similarly for $\Psi^{I}, \Psi^{J}$. This complexification is, in fact, crucial for the consistency of our asymmetric construction and it will permit us to represent the partition function of the theory in terms of simple level-1 characters.

In terms of the fermionization fields $Y^I,W^I$, we can represent the orbifold action $P(\epsilon)$ on the $T^5$-coordinates as:
\begin{align}\nonumber
	& Y_L^I \,\rightarrow\, {(UPU^{-1})^I}_J(\epsilon) Y_L^J=\lambda_I(\epsilon)\, Y_L^I~,\\
	& W_L^I \rightarrow {(UPU^{-1})^I}_J(1) W_L^J = \lambda_I(1)\, W_L^I~.
\end{align}
It is important to note that the $Z^I$ are not fermionized into the simple product $Y^I W^I$ but, rather, into a linear combination of fermion bilinears:
\begin{align}\nonumber
	 i\partial Z^I &= i\sum\limits_{J=1}^{N}{\,{U^I}_J(\epsilon)\,{(U^{-1})^J}_K(\epsilon)\,{(U^{-1})^J}_L(1)\,Y^K\,W^L}~\\
	& = \frac{i}{\sqrt{N}}\,\sum\limits_{K,L=1}^{N}{\delta_{(I-K-L+1)\,{\textrm{mod}}\,N,0}\,Y^K W^L}~,
\end{align}
where the Kronecker $\delta$-function restricts the sum to those $K,L$ which satisfy the constraint $(I-K-L+1)\,{\textrm{mod}}\,N=0$. This constraint picks up precisely the combinations that correctly reproduce the $\mathbb{Z}_N$-transformation of $\partial Z^I$, as expected.

The above fermionization of the coordinates is only consistent at the fermionic point in moduli space. In our conventions, this is a square lattice $G_{IJ}=r^2\delta_{IJ}$, with $r=1/\sqrt{2}$ being the fermionic radius. The orbifold action typically introduces additional constraints. First of all, a non-trivial requirement is that the orbifold action $P(\epsilon)$ must be a symmetry of the local $\mathcal{N}=1$ superconformal theory (SCFT) on the worldsheet.
It is straightforward to check that both $T_B$ and $T_F$ of the internal SCFT are invariant, since at the fermionic point the orbifold acts crystallographically:
\begin{align}\nonumber
	& T^{\textrm{int}}(z) = -r^2\,\sum\limits_{I=1}^{N}{(\partial X^I)^2}-\frac{r^2}{2}\,\sum\limits_{I=1}^{N}{\psi^I_L\partial \psi^I_L}+\ldots = -r^2~\widehat{\sum\limits_{I,J}}{\left(\partial Z^I \partial Z^J -\frac{1}{2}\,\Psi_L^I \partial \Psi_L^J\right)}+\ldots,\\
	& T_F^{\textrm{int}}(z) = i\sqrt{2}\,r^2\,\sum\limits_{I=1}^{N}{\psi_L^I\, \partial X^I}+\ldots=  i\sqrt{2}\, r^2~\widehat{\sum\limits_{I,J}}{\,\Psi_L^I\,\partial Z^J}+\ldots~,
\end{align}
where  $\widehat{\sum}$ stands for the sum subject to the constraint $I+J\in N\mathbb{Z}+2(1-\nu)$ and the dots denote contributions of the remaining coordinates in $T^5$ that are not transformed, as well as the contributions due to the (super-)ghosts and the $S^1$ (super-)coordinate. Of course, at the fermionic point, $T_B$ and $T_F$ can be realized entirely in terms of the free fermions $\psi^I,y^I$ and $\omega^I$ and similar relations hold.

In particular, the correct implementation of the orbifold projections require the presence of well-defined $R$-symmetry charges, $J=\Psi\bar\Psi$, reflecting the mutual locality of OPEs between vertex operators in the left-moving CFT. Hence, this implies that the free fermions $\{\Psi^I, Y^I, W^I\}$ can always be grouped into complex conjugate pairs, so that the currents\footnote{Note that   real $Y$ and $W$ fermions that transform with the same real eigenvalue ($\lambda=\pm 1$) under the orbifold, can also form `mixed' invariant currents of the form $J_{YW}(z)=YW$.}:
\begin{align}\nonumber
	& J_\Psi(z)= \Psi \bar\Psi~, \\ \nonumber
	& J_Y(z)= Y\bar Y~,\\
	& J_W(z) = W\bar W~,
\end{align}
remain invariant under the orbifold action. Obviously, this is always the case for the fermions $\Psi^I, Y^I, W^I$ corresponding to complex eigenvalues $\lambda_I$. The case of fermions corresponding to real eigenvalues $\lambda=\pm 1$, however, is more subtle. Unless the fermions transforming with $\lambda=-1$ can be grouped into complex pairs to form invariant currents, the fermionization will lead to non-local OPEs in the CFT and to the breakdown of the modular invariance of the theory. This is, essentially, equivalent to the requirement of a consistent particle interpretation (unitarity) \cite{Narain:1986qm}.

After this general discussion, we can now return to the case of $T^5\times S^1$ and identify the possible consistent asymmetric orbifolds that may be obtained from our construction. We start with the pure permutations ($\epsilon=+1$) and note that for odd $N$ (corresponding to $\mathbb{Z}_3$ and $\mathbb{Z}_5$ acting on $T^3$ and $T^5$, respectively), all eigenvalues except for the identity are complex. For even $N$ (acting on $T^N$), however, there is always a single $\lambda=-1$ eigenvalue. In the case of $\mathbb{Z}_2$, one may extend the action over $T^4\subset T^5$, simply by defining the $\mathbb{Z}_2$-permutation directly on the complexified (super-)coordinates of $T^4$.

In the case of permutation with reflection ($\epsilon=-1$) acting on $T^N$, there arises a subtlety due to the fact that the $\Psi^I,Y^I$-fermions are permuted with reflection (according to $P(-1)$), whereas the $W^I$-fermions only transform as pure permutations (according to $P(+1)$). As a result, if the orbifold action is restricted on $T^N$, there will always be fermions transforming with $\lambda=-1$ that cannot be complexified. The only way out is, again, to extend the orbifold action to $T^{2N}$ and define the permutation (with reflection) $P$ directly on the complexified (super-)coordinates. The relevant case, here, is the $N=2$ defined on $T^4$, which acts as a  $\mathbb{Z}_4$-rotation on each of the two $T^2$-subspaces.

Of course, one could have considered more general asymmetric orbifolds\footnote{Examples of other constructions can be found, for instance, in \cite{Angelantonj:2000xf,Blumenhagen:2000fp}.}. However, given our ansatz that the orbifold only acts on the left-moving part of a $T^5$ at the fermionic point, we are essentially restricted to the four cases: $\mathbb{Z}_2, \mathbb{Z}_3, \mathbb{Z}_4$ and $\mathbb{Z}_5$ described above.


\subsection{Partition functions}\label{PartitionFunctionsSection}
\bigskip

Up to now we have been discussing the asymmetric $\mathbb{Z}_N$ action on the left-moving $T^5$ coordinates. In order to make the action of the orbifold free, that is, in order to ensure the absence of fixed points, we need to entangle the asymmetric rotation of the $T^5$ with a translation along the $S^1$-direction as in (\ref{GeneralOrbifold}). The resulting orbifold will not contain twisted sectors nor will it enforce any projections to invariant states. Rather, it will have the effect of shifting the masses of states `charged' under the $\mathbb{Z}_N$-rotation.

We will work in the Scherk-Schwarz basis and start by writing the boundary conditions of the $S^1$-coordinate along the two non-contractible cycles of the worldsheet torus as:
\begin{align}\nonumber
	& \mathbb{X}^a(\sigma^1+2\pi,\sigma^2) = \mathbb{X}^a(\sigma^1,\sigma^2)+2\pi n^a~,\\
	& \mathbb{X}^a(\sigma^1,\sigma^2+2\pi) = \mathbb{X}^a(\sigma^1,\sigma^2)+2\pi \tilde{m}^a~.
\end{align}
Before we define the torus partition function of the theory, it will be convenient to define the orbifold twist variables:
\begin{align}
	h = \frac{2n}{N}~~,~~g = \frac{2\tilde{m}}{N}~,
\end{align}
where $\tilde{m},n\in\mathbb{Z}$ are the winding quantum numbers around the $S^1$, as each of the worldsheet coordinates $\sigma^1,\sigma^2$ encircle the two non-trivial cycles of the worldsheet torus.

We can now write the modular invariant partition function of the model (up to an overall $\tau_2$-power) in the following form:
\begin{align}\label{GeneralPartition}
	Z = \frac{1}{\eta^{12}\,\bar\eta^{12}}~R\sum\limits_{\tilde{m},n\in\mathbb{Z}}{e^{-\frac{\pi R^2}{\tau_2}|\tilde{m}+\tau n|^2}~ Z_L[^h_g](\tau)\tilde{Z}_R(\bar{\tau})\,\Gamma_{(5,5)}[^h_g](\tau,\bar\tau) }~.
\end{align}
Here, $Z_R[^h_g]$ and $\tilde{Z}_R$ are the contributions of the left- and right- moving worldsheet fermion superpartners $\psi,\tilde{\psi}$, respectively and $\Gamma_{(5,5)}[^h_g]$ is the $(5,5)$-lattice associated to the $T^5$. The exponential factor in (\ref{GeneralPartition}) is the Lagrangian representation of the lattice partition function of the $S^1$ coordinate, $\mathbb{X}$.

Let us analyze each of the above pieces, starting with the partition function $Z_L[^h_g]$ which contains the sum over the spin-structures of  the left-moving worldsheet fermions. The boundary conditions of the real worldsheet fermions in the original basis now act as the repeated permutations to the power dictated by the $(\tilde{m},n)$-winding numbers around the $S^1$:
\begin{align}\nonumber
	& \psi^I(\sigma^1+2\pi,\sigma^2) = {(P^n)^I}_J~e^{i\pi(1-a)}\,\psi^J(\sigma^1,\sigma^2)~,\\ \label{BCsfermion}
	& \psi^I(\sigma^1,\sigma^2+2\pi)= {(P^{\tilde m})^I}_J~e^{i\pi(1-b)}\,\psi^J(\sigma^1,\sigma^2)~.
\end{align}
As pointed out at the end of the previous section, the consistency of our construction requires that the worldsheet fermions in the diagonal $\Psi_L^I$-basis can always be separated into complex conjugate pairs $(\Psi_L, \bar\Psi_L)$ with the same spin structures and into real fermions that are invariant under the orbifold action. Assuming that a complex conjugate fermion pair $f=(\Psi_L^{(f)}, \bar\Psi_L^{(f)})$ transforms with complex phase  $\lambda_f$:
\begin{align}
	\left.\begin{array}{l l}
		\Psi_L^{(f)} \rightarrow \lambda_f ~\Psi_L^{(f)} \\
		\bar\Psi_L^{(f)} \rightarrow \lambda^*_f ~ \bar\Psi_L^{(f)} \\
	\end{array}\right\} \quad , \quad
\textrm{with} \quad
			\lambda_f = e^{\frac{2\pi i}{N} q_f} ~,
\end{align}
 we can express $Z_L[^h_g]$ as:
\begin{align}
	Z_f[^h_g] = \frac{1}{2}\sum\limits_{a,b=0}^{1}{(-)^{a+b}~\theta[^a_b]^{4-N_f}\left(\,\prod\limits_{f-\textrm{complex}}{ \theta\left[^{a-q_f h}_{b-q_f g}\right]} e^{-i\pi\left[q_f^2\frac{hg}{4}-b q_f\frac{h}{2}\right]}\right) }~,
\end{align}
where $N_f$ is the number of complex $\Psi_L$-pairs. This expression is written with the understanding that complex conjugate pairs can always be grouped together into single $\theta$-functions (of integral power) describing each conjugate fermion pair, consistently with the requirements of modular invariance. Notice the presence of the phase accompanying each $\theta$-factor in the product. This phase arises naturally as the unique (continuous) deformation of the theory that preserves the modular properties. It can be seen as the `back-reaction' of the theory in order to preserve modular invariance \cite{Ferrara:1987qp,Kounnas:1989dk,Kiritsis:1997ca}.

The contribution of the right-moving worldsheet fermions is that of the usual, undeformed Type II theory:
\begin{align}
	\tilde{Z}_{R}= \frac{1}{2}\sum\limits_{\bar{a},\bar{b}=0}^{1}{(-)^{\bar{a}+\bar{b}+\mu\bar{a}\bar{b}}~\bar\theta[^{\bar{a}}_{\bar{b}}]^4} ~,
\end{align}
with $\mu=0,1$ is a chirality choice, distinguishing between Type IIB and IIA theories, respectively.

We now turn to the bosonic $T^5$ coordinates. In the original basis, the boundary conditions of the bosons include the same permutations as their worldsheet superpartners, given in (\ref{BCsfermion}):
\begin{align}\nonumber
	&  X^I(\sigma^1+2\pi,\sigma^2) = {(P^n)^I}_J\, X^J(\sigma^1,\sigma^2)~,\\
	&  X^I(\sigma^1,\sigma^2+2\pi)= {(P^{\tilde{m}})^I}_J\, X^J(\sigma^1,\sigma^2)~.
\end{align}
Since we are working at the fermionic point, the partition function of the `twisted' $(5,5)$-lattice can be written entirely in terms of level-1 $\theta$-characters. As before, we assume that complex fermion pairs $f=(Y_L,\bar{Y}_L)$ and $f'=(W_L,\bar{W}_L)$ transform with the complex phases $\lambda_f = e^{2\pi i q_f/N}$ and $\lambda_{f'}=e^{2\pi i q_{f'}/N}$, respectively.
The fermionic lattice partition function is then written as:
\begin{align}\nonumber
	\Gamma_{(5,5)}[^h_g] =& \,\frac{1}{2}\sum\limits_{\gamma,\delta=0}^{1} \theta[^\gamma_\delta]^{5-N_f-N_{f'}}~\bar\theta[^\gamma_\delta]^5 \left( \prod\limits_{f-\textrm{complex}} \theta\left[^{\gamma-q_f h}_{\delta-q_f g}\right] e^{-i\pi\left[q_f^2\frac{hg}{4}-\delta q_f\frac{h}{2}\right]} \right) \\ \label{Partition3}
 &\times \left(\prod\limits_{f'-\textrm{complex}} \theta\left[^{\gamma-q_{f'}h}_{\delta-q_{f'}g}\right] e^{-i\pi\left[q_{f'}^2\frac{hg}{4}-\delta q_{f'}\frac{h}{2}\right]}\right)~,
\end{align}
where the integers $N_f, N_{f'}$ stand for the total numbers of complex conjugate $(Y_L,\bar{Y}_L)$ and $(W_L,\bar{W}_L)$ fermion pairs, respectively. Hence, the $\theta$-functions outside the parentheses correspond to the invariant left- and right-moving coordinates, whereas the parentheses in the first and second line are the contributions of the twisted left-moving auxiliary fermions $Y_L$ and $W_L$, respectively. Again, in all consistent orbifold models, all free fermions will be organized into pairs sharing the same spin structures, according to their eigenvalues, in order to form integral-powered $\theta$-characters.

It is straightforward to check that the full partition function (\ref{GeneralPartition}) is indeed modular invariant, up to an overall $\tau_2$-factor. Moreover, the partition function can be seen to arise as the modular-invariance preserving deformation of the ordinary toroidal compactification on $T^5\times S^1$. It is compatible with the constraints arising from higher-genus modular invariance and factorization and, hence, defines a class of consistent orbifold vacua.

Finally, before we proceed further, it is instructive to make a comment on the form of the partition function presented above. In particular, expressed as in (\ref{GeneralPartition}), the partition function exhibits manifestly the correlation of the fermion spin structures $[^a_b]$, $[^\gamma_\delta]$ with the winding numbers $\tilde{m},n$ around the $S^1$ or, in other words, the correlation between the $S^1$ winding numbers and the (asymmetric) boundary conditions of the (super-)coordinates $Z^I,\Psi^I$ of $T^5$. This representation in which the $S^1$-lattice appears in its Lagrangian form is useful in making contact with the (fermionic) $\sigma$-model, as it is in this form that the partition function arises from the 1-loop path integral on the worldsheet torus. As we will discuss in the Section \ref{ScherkSchwarzSection}, it is in this picture that the freely-acting orbifold can be most naturally re-interpreted as a stringy Scherk-Schwarz flux.

It is straightforward to recast (\ref{GeneralPartition}) into its Hamiltonian representation in the orbifold basis, which is the natural representation that arises if one considers the trace over the Hilbert space. This requires the redefinition of the winding summation variables as:
\begin{align}\nonumber
	& n= N\hat{n}+H~,\\
	& \tilde{m}= N\hat{\tilde{m}}+G~,
\end{align}
where $\hat{n},\hat{\tilde{m}}$ are unconstrained windings and there is an independent summation over $H,G\in\mathbb{Z}_N$. Using the periodicity properties of Jacobi $\theta$-functions, one may eliminate the winding dependence from the $\theta$-characters, giving rise to additional phases. Finally, a Poisson resummation in $\hat{\tilde{m}}$ resets the $S^1$-lattice into its Hamiltonian representation and the sum over $H,G\in\mathbb{Z}_N$ now plays the r\^{o}le of the orbifold summation over `twisted' sectors and their corresponding `projections', whereas the resulting unconstrained summation variables $\hat{m},\hat{n}$ are now naturally interpreted as the momentum and winding quantum numbers around the $S^1$, respectively. The radius of the $S^1$ in the orbifold representation, $R_{\textrm{orb}}$ can be easily related to the corresponding radius $R$ in the Scherk-Schwarz picture, through $R_{\textrm{orb}}=N R$. This completes the equivalence between the two representations.



\section{Non-commutativity from Non-Geometric Fluxes}\label{SectionNonCommut}
\bigskip

So far, we have reviewed the construction of consistent freely-acting asymmetric $\mathbb{Z}_N$-orbifolds.
As discussed in Section \ref{SectionMotivation}, one has to keep in mind that these backgrounds are to be considered as (smooth) orbifold limits of string compactifications with  non-geometric $Q$-fluxes (T-folds).
We are now ready to examine the algebra of commutators of the internal coordinates, subject to the $\mathbb{Z}_N$-orbifold boundary conditions (\ref{ZNbc}). We will see that the asymmetric nature of the orbifold leads to a non-commutative  algebra for the  $T^N$ coordinates, the structure constants of which are nicely parametrized in terms of an asymmetric Scherk-Schwarz flux matrix ${F^I}_J$.

\subsection{Non-commutative algebra}
\bigskip

We will start by writing down the mode expansions directly in the basis $Z^I$ which diagonalizes the orbifold action. We restrict our attention to the left-movers:
\begin{align}
	Z^I_L(\tau,\sigma)= \frac{i}{\sqrt{2}}\sum\limits_{k\in\mathbb{Z}}{\frac{\alpha_{k-\theta^I}^I}{k-\theta^I}~e^{-i(k-\theta^I)(\tau+\sigma)}}~,
\end{align}
where the modes satisfy the standard commutation relations:
\begin{align}\label{Acommut}
	[\alpha^I_{k-\theta^I},\alpha^J_{\ell-\theta^J}] = (k-\theta^I)\delta_{k+\ell,0}\,(UG^{-1}U^T)^{IJ}~.
\end{align}
Notice that:
\begin{align}\label{newMetric}
	(UG^{-1}U^T)^{IJ} = \frac{1}{r^2}\,\delta_{[I+J-2(1-\nu)]\,\textrm{mod}\, N,0}~,
\end{align}
 is nothing but the inverse metric in the $Z^I$-basis.

It is convenient to parametrize the orbifold action ${P^I}_J(\epsilon)$ in terms of a real antisymmetric matrix ${F^I}_J$ :
\begin{align}
		{P^I}_J={\left(e^{2\pi  F }\right)^I}_J~.
\end{align}
The reason is the following. As we will see in Section \ref{ScherkSchwarzSection}, the freely-acting asymmetric orbifolds we considered admit a natural description in terms of compactifications on non-trivial backgrounds with constant asymmetric Scherk-Schwarz flux, which will be precisely identified with ${F^I}_J$. One may keep in mind, however, that the flux matrix ${F^I}_J$ is a different manifestation of the $Q$-flux, at the special orbifold point where the CFT is exactly solvable.

 The deformation angle $\theta^I=f^I n$ is nothing but the $F$-flux eigenvalue $f^I$ times the winding number along $S^1$ and is  related to the eigenvalue $\lambda_I$ through:
 \begin{align}
 	e^{2\pi i \theta^I} = \lambda_I^n = e^{2\pi i f^I n}~.
 \end{align}
Explicitly, the matrix ${F^I}_J$ can be easily constructed from its eigenvalues, $\pm i f^I$:
\begin{align}\label{Feigenvalues}
	  f^I = \frac{(I-1+\nu)}{N} \quad , \quad \textrm{with}\quad I=1, \ldots,\left[\tfrac{N+1}{2}\right]~,
\end{align}
and is given by the following expression:
\begin{align}\label{FQuantiz}
	{F^I}_J = -\frac{2}{N}\sum\limits_{K=1}^{\left[\frac{N+1}{2}\right]}{f^K\,\sin\left(2\pi f^K (I-J)\right)} ~,
\end{align}
where the square brackets in the upper limit of the sum denote the integer part.

 The commutation relations (\ref{Acommut}) should be supplemented with the `reality condition':
\begin{align}\label{realityCond}
	 \left.
	\begin{array}{l}
	 \overline{\alpha^I}_{k-\theta^I} = \alpha^{J}_{-k+\theta^I}   \\
	  \theta^I= -\theta^J \\
	  \end{array}
	  \right\}  \quad ,\quad \textrm{if}~~\overline{Z^I}=Z^J~,
\end{align}
which essentially reflects the fact that only commutators between complex conjugate pairs are non-vanishing.

We can now take a pair of coordinates and evaluate their equal-times commutator using  (\ref{Acommut}), (\ref{newMetric}), and (\ref{realityCond}):
\begin{align}
	[Z^I_L(\tau,\sigma),Z^J_L(\tau,\sigma')] = \frac{1}{2}(UG^{-1}U^T)^{IJ}\, \sum\limits_{k\in\mathbb{Z}}{\frac{e^{-i(k-\theta^I)(\sigma-\sigma')}}{k-\theta^I}}~.
\end{align}
Setting $z=e^{i(\sigma'-\sigma)}$, we can express the series as the linear combination of hypergeometric functions in (\ref{hypergeometric}), which can then be analytically continued to $z\rightarrow 1$ :
\begin{align}\label{Theta}
	\Theta(\theta^I) = \Bigg\{ \begin{array}{c l}
						 -\pi \cot(\pi \theta^I) & \quad,\quad \theta^I \notin \mathbb{Z} \\
						 0 &  \quad,\quad  \theta^I \in \mathbb{Z} \\
						\end{array}~.
\end{align}
Going back to the original basis, the commutation relations between the full coordinates $X=X_L+X_R$ become:
\begin{align}\label{GeneralAlgebra}
	[X^I,X^J]\Big|_{\sigma=\sigma'} = \frac{1}{2}\,\Theta(in F)^{IJ}~.
\end{align}
Note that the argument of the function $\Theta$ in (\ref{GeneralAlgebra}) is now a matrix and the easiest way to define it is through its eigenvalues. Since $F$ is at most five-dimensional and antisymmetric, it can always be expanded as:
\begin{align}\label{ThetaExp}
	\Theta(in F) = \alpha F + \beta F^3~,
\end{align}
with $\alpha,\beta$ complex coefficients that are determined by substituting in the eigenvalues of $F$, eq. (\ref{Feigenvalues}).

In the cases $\mathbb{Z}_2,\mathbb{Z}_3$ and $\mathbb{Z}_4$, there is only a single complex eigenvalue\footnote{In the $\mathbb{Z}_4$ case with $N=2$, $\epsilon=-1$, the complex eigenvalue comes with multiplicity two because of the doubling of the orbifold action on $T^4$.} and, hence, $\beta=0$. For the $\mathbb{Z}_k$ orbifold with $k=2,3,4$, the non-commutative algebra between the toroidal coordinates then takes the simple form:
\begin{align}\label{Z234commut}
	 [X^I,X^J] = \tfrac{i}{2} k\, F^{IJ}\,\Theta\left( n/k\right) = \left\{ \begin{array}{c l}
	 						 -\tfrac{i}{2} \pi k\, F^{IJ}\,\cot\left(\frac{\pi n}{k}\right)  & ,\quad n \notin k\mathbb{Z} \\
							 0 & ,\quad n \in  k\mathbb{Z} \\
	 				\end{array}
	 \right.	 ~.
\end{align}
The case of $\mathbb{Z}_2$ is special, because the r.h.s. vanishes identically and the toroidal coordinates remain commutative. This is, essentially, due to the fact that the $\mathbb{Z}_2$-action can always be reduced to a reflection of a real coordinate and, hence, it does not really `entangle' different coordinates.

Notice that in (\ref{Z234commut}), the structure constants of the algebra are proportional to the flux $F^{IJ}$ itself. This simple behavior is not, however, true in general. In particular, the $\mathbb{Z}_5$-case has richer structure, because ${F^I}_J$  then has two distinct complex eigenvalues and $\alpha,\beta$ are given as the linear combinations:
\begin{align}\nonumber
	&\alpha = \tfrac{5}{6} i\Big[8\,\Theta(n/5)-\Theta(2 n/5)\Big] ~,\\
	&\beta = \tfrac{125}{6} i\Big[2\,\Theta( n/5)-\Theta(2 n/5) \Big] ~.
\end{align}
There is, hence, a second contribution to the structure constants (\ref{ThetaExp}) that is qubic in the flux.

Before proceeding to analyze particular examples, it is instructive to comment on the structure of the general result (\ref{GeneralAlgebra}). First of all, the stringy origin of the non-commutative behavior becomes manifest, by noticing that $[X^I,X^J]$  is always proportional to the string scale $\alpha'$. Secondly, the explicit periodic dependence on the winding number $n$ along the circle direction $\mathbb{X}$, implies that only when non-trivial winding modes along $S^1$ can be excited, can one probe the non-commutativity in the toroidal coordinates. This is consistent with the fact that, in the limit where the radius of $S^1$ becomes large $R/\alpha' \gg 1$ and states carrying non-trivial winding charge become supermassive, the fibration of $T^2$ over $S^1$ trivializes and the non-commutative effects are screened.


\subsection{An Explicit Example: the asymmetric $\mathbb{Z}_3$}
\bigskip

Having discussed how non-geometric Scherk-Schwarz $F$-flux leads to a non-commutative algebra in the internal space coordinates, we will now pick  $\mathbb{Z}_3$ as a particular toy example and present the explicit construction of the orbifold and its non-commutative algebra in some detail.

As discussed in Section \ref{ZNOrbifoldConstruction}, the  orbifold acts on a $T^3\subset T^5$ torus as the pure permutation, $P_{IJ}(1)$, defined in (\ref{OrbifoldAction}). Its eigenvalues are:
\begin{align}
	\lambda_1=1 \quad,\quad \lambda_2=\overline{\lambda_3}=e^{2\pi i/3}~.
\end{align}
We then define the basis $Z^I$ that diagonalizes the orbifold action:
\begin{align}
\begin{split}
	& Z^1 = \tfrac{1}{\sqrt{3}}\left(X^1+X^2+X^3\right)~,\\
	& Z^2 = \tfrac{1}{\sqrt{3}}\left(X^1+e^{2\pi i/3} X^2 + e^{-2\pi i/3} X^3\right)~,\\
	& Z^3 = \overline{Z^2}~,
\end{split}
\end{align}
and similarly for the fermions $\Psi^I={U^I}_J(1)\psi^J$. We, subsequently, fermionize the bosons $\partial X^I=y^I\omega^I$ and introduce the new bases $Y^I={U^I}_J(1)\psi^I$, $W^I={U^I}_J(1)\omega^J$. The bosonic currents are then expressed as linear combinations of free bifermion currents:
\begin{align}
\begin{split}
	i\partial Z^1 = \tfrac{i}{\sqrt{3}}\left(Y^1 W^1+Y^2 W^3+Y^3 W^2\right)~,\\
	i\partial Z^2 = \tfrac{i}{\sqrt{3}}\left(Y^1 W^2+Y^2 W^1+Y^3 W^3\right)~,\\
	i\partial Z^3 = \tfrac{i}{\sqrt{3}}\left(Y^1 W^3+Y^2 W^2+Y^3 W^1\right)~.
\end{split}
\end{align}
Furthermore, the internal energy-momentum tensor $T_B$ and the worldsheet supercurrent $T_F$ close into a consistent $\mathbb{Z}_3$-invariant $\mathcal{N}=1$ SCFT, realized entirely in terms of free fermions:
\begin{align}
\begin{split}
	T_B(z)=\tfrac{r^2}{2}\Big[ &\Psi^1\partial \Psi^1+\Psi^2 \partial\Psi^3+\Psi^3\partial \Psi^2+Y^1\partial Y^1+W^1\partial W^1 +Y^2\partial Y^3+Y^3\partial Y^2\Big]+\ldots \\
	 T_F(z)= \tfrac{ir^2}{\sqrt{6}}\Big[&\Psi^1(Y^1 W^1+Y^2 W^3+Y^3 W^2)  \\
	+&\Psi^2(Y^1 W^2+Y^2 W^1+Y^3 W^3)+\Psi^3(Y^1 W^3+Y^2 W^2+Y^3 W^1)\Big]+\ldots
\end{split}
\end{align}
The modular invariant\footnote{Up to irrelevant $\tau_2$-factors.} partition function of the model can be easily obtained by working along the lines of Section \ref{PartitionFunctionsSection}:
\begin{align}
	Z=\frac{R}{\eta^{12}\bar\eta^{12}}\sum\limits_{\tilde{m},n\in\mathbb{Z}}{e^{-\frac{\pi R^2}{\tau_2}|\tilde{m}+\tau n|^2}}\Bigg[\tfrac{1}{2}\sum\limits_{a,b=0}^{1}{(-)^{a+b}\theta[^a_b]^3 \theta[^{a-h}_{b-g}]}~\tfrac{1}{2}\sum\limits_{\bar{a},\bar{b}=0}^{1}{(-)^{\bar{a}+\bar{b}+\mu\bar{a}\bar{b}}\bar\theta[^{\bar{a}}_{\bar{b}}]^4}\Gamma_{(5,5)}[^h_g]\,e^{-i\pi \Phi}\Bigg]~,
\end{align}
where $h=\frac{2}{3}n$,  $g=\frac{2}{3}\tilde{m}$ and the $(5,5)$-lattice is expressed in terms of $\theta$-characters as:
\begin{align}
	\Gamma_{(5,5)}[^h_g] = \tfrac{1}{2}\sum\limits_{\gamma,\delta=0}^{1}{\theta[^\gamma_\delta]^3 \theta[^{\gamma-h}_{\delta-g}]^2 ~\bar\theta[^\gamma_\delta]^5}~.
\end{align}
The `modular-balancing' phase is given in terms of the orbifold parameters $h,g$ and the fermionic spin-structures:
\begin{align}
	\Phi = \tfrac{3}{4}hg-\tfrac{1}{2}(b+2\delta)h~.
\end{align}
The freely-acting asymmetric $\mathbb{Z}_3$ orbifold model we have constructed can be now reinterpreted as a compactification on $T^5\times S^1$ in the presence of a non-geometric flux:
\begin{align}
	{F^I}_J = \frac{1}{3\sqrt{3}}\left(\begin{array}{r r r}
								0 & 1 & -1 \\
								-1 & 0 & 1 \\
								1& -1 & 0 \\			
							\end{array}\right)~,
\end{align}
with eigenvalues $f^1=0$, $f^2=1/3$ and $f^3=-1/3$.

The only non-trivial commutation relation in the $Z^I$-basis is between the complex conjugate pair:
\begin{align}
	[Z^2,Z^3] = \frac{1}{2r^2}\Theta(n/3)~,
\end{align}
where $\Theta(x)$ was defined in (\ref{Theta}). Translating this result in the original $X^I$-basis, one readily obtains:
\begin{align}
	[X^I,X^J] = \tfrac{3}{2} i F^{IJ} \Theta(n/3)~.
\end{align}
Picking, for example, a  state with unit winding $n=1$ and reinstating the explicit $\alpha'$-dependence, one can express the non-vanishing commutators as:
\begin{align}
	[X^I,X^J] = \frac{i\pi\alpha'}{3}\left(\begin{array}{r r r}
								0 & -1 & 1 \\
								1 & 0 & -1 \\
								-1& 1 & 0 \\			
							\end{array}\right)^{IJ} ~,
\end{align}
with $I,J$ running in the $T^3$-directions, parametrized by $X^1,X^2,X^3$.


\section{Stringy Scherk-Schwarz and Non-Geometric Fluxes}\label{ScherkSchwarzSection}
\bigskip

So far, we have reviewed the construction of asymmetric freely-acting $\mathbb{Z}_N$-orbifolds acting as permutations with possible reflections on the left-moving $T^5$-coordinates, accompanied by a shift in both the momenta and windings of an $S^1$. Furthermore, we have shown that the toroidal coordinates become non-commutative due to the asymmetric nature of the orbifold. It is important to note that, in the orbifold picture, the presence of the $Q$-flux is essentially hidden in the non-geometric monodromy of the worldsheet fields. In this section we will reinterpret this construction by performing a field redefinition that absorbs the non-trivial boundary conditions of the worldsheet fields and translates this asymmetric monodromy ($Q$-flux) into a non-trivial $\sigma$-model flux background. This procedure will be identified with an asymmetric version of the stringy Scherk-Schwarz mechanism, realized by turning on a constant non-geometric flux associated to the $U(1)_R$ gauge group of the circle, along the directions of $T^5$.


\subsection{The Symmetric Scherk-Schwarz}
\bigskip

Before dealing with the asymmetric case, it will be instructive to first review the stringy realization of the conventional `symmetric' Scherk-Schwarz mechanism \cite{Ferrara:1987qp,Kounnas:1989dk,Kiritsis:1997ca}. We will start from the original $\mathcal{N}_4=8$ Type II theory on $T^5\times S^1$, described by the internal $\sigma$-model:
\begin{align}
	S_0 = \frac{1}{4\pi}\int{d^2 z \left(2G_{IJ}\partial \hat{X}^I\bar\partial \hat{X}^J+G_{IJ}\hat\psi^I\bar\partial\hat\psi^J\right)}+\ldots \quad,
\end{align}
and turn on the following deformation:
\begin{align}\label{Deformation}
	 S_{\textrm{def}} = -i\int{d^2 z~F_{IJ}^{a}\left[(\hat\psi^I\hat\psi^J-{\hat X^I\stackrel{\leftrightarrow}{\partial} \hat X^J})\,\tilde{J}^a+(\hat{\tilde{\psi}}^I \hat{\tilde{\psi}}^J-{\hat{X}^I\stackrel{\leftrightarrow}{\bar\partial} \hat{X}^J})\, J^a\right]  }~,
\end{align}
where $F_{IJ}^a$ is a constant five-dimensional antisymmetric matrix in $I,J=1,\ldots,N\leq 5$, valued in the Lie algebra of the $U(1)_{L,R}$ gauge groups generated by the left- and right-moving $(1,0)$ and $(0,1)$-currents $J^a$ and $\tilde{J}^a$, respectively. In our case, $J^a$ and $\tilde{J}^a$ will be identified with the $U(1)$ currents arising from the circle compactification of $\mathbb{X}$ on $S^1$, namely, $J^{\mathbb{X}}(z)=i\partial \mathbb{X}(z)$ and $\tilde{J}^\mathbb{X}(\bar{z})=i\bar\partial \mathbb{X}(\bar{z})$, respectively. Notice that the particular form of the perturbation is consistent with the local $\mathcal{N}=(1,1)$ worldsheet superconformal algebra.

The deformation (\ref{Deformation}) describes a constant field strength $F_{IJ}^a$ for the $U(1)$ associated to the circle, in the directions $I,J$. This can be seen from the bosonic part, $F_{IJ}^a\,\hat{X}^I\partial \hat{X}^J\,\tilde{J}^a$, of the perturbation by identifying:
\begin{align}
	A_I^a(\hat{X})= -F_{IJ}^a\,\hat{X}^J~,
\end{align}
as the associated gauge field. In terms of $A^a_I$, the bosonic part of the perturbation can be rewritten as:
\begin{align}
	2i\int{d^2 z~A_I^a(\hat{X})~\partial \hat{X}^I\, \tilde{J}^a}~,
\end{align}
which is precisely the $\sigma$-model perturbation describing a non-trivial gauge field background in the zero ghost picture.

From the higher-dimensional point of view, the $F_{IJ}^a$ deformations (\ref{Deformation}) have the interpretation of non-trivial $F$-flux backgrounds on the $T^N$, whereas from the four-dimensional point of view they can be regarded as non-trivial vacuum expectation values for auxiliary fields in the gauged supergravity description.

Notice that (\ref{Deformation}) is not a marginal deformation for generic values of $F_{IJ}^a$. This can be seen, for example, from the fact that $\hat{X}^I\partial \hat{X}^J$ is not a well-defined $(1,0)$ conformal operator. However, for special quantized values of $F_{IJ}^a$ the operator:
\begin{align}\label{Rotation}
	\exp\int{F_{IJ}^a\,(\hat\psi^I\hat\psi^J-{\hat X^I\stackrel{\leftrightarrow}{\partial} \hat X^J})}~,
\end{align}
and its right-moving analogue become consistent $O(5,5;\mathbb{Z})$ rotation operators, acting crystallographically as automorphisms of the $R$-symmetry and $T^5$ lattices. It is for these quantized values of $F_{IJ}^a$ for which the operator (\ref{Rotation}) commutes with the worldsheet supercurrent $T_F(z)$ that the deformation (\ref{Deformation}) becomes integrable and the CFT remains solvable.

Let us briefly illustrate how this works by focusing on the left-moving bosonic term of (\ref{Deformation}). Our aim is to perform a field redefinition $\hat{X}^I \rightarrow X^I$ in the bosons that absorbs the bosonic part of the deformation into the kinetic term of free scalars:
\begin{align}
	2G_{IJ}\partial X^I\bar\partial X^J~.
\end{align}
 The natural choice is to introduce new (unhatted) fields $X^I$, defined as:
\begin{align}\label{FieldRedefinitionBoson}
	X^I \equiv{\left(e^{F^a \mathbb{X}^a}\right)^I}_J \,\hat X^J~.
\end{align}
However, this field redefinition can only absorb (the bosonic part of) the perturbation of the $\sigma$-model considered above into free kinetic terms  at the cost of introducing a non-linear backreaction to the metric element $G_{\mathbb{XX}}$, associated to the $S^1$. This, however, can be taken care of from the beginning, by introducing an additional correction term to our perturbation in order to preserve the flatness of the resulting CFT.

The corresponding field redefinition for the worldsheet fermions is more subtle, since we need to respect the $\mathcal{N}=1$ worldsheet SCFT. The correct form can be easily constructed by working in a manifest $\mathcal{N}=1$ worldsheet superspace formalism and leads to the definition of new (unhatted) fermionic fields $\psi^I$: 
\begin{align}\label{FieldRedefinition}
	\psi^I \equiv{\left(e^{F^a \mathbb{X}^a}\right)^I}_J \left[ \hat{\psi}^J + {F^J}_K \hat{X}^K\psi^{\mathbb{X}}\right]~,
\end{align}
with $\psi^{\mathbb{X}}$ being the fermionic superpartner of $\mathbb{X}$ and with $F^a \mathbb{X}^a$ being the matrix ${(F^a)^I}_J\,\mathbb{X}^a$, preserving the metric:
\begin{align}
	\left(e^{-F^a \mathbb{X}^a}\right)^T G\, e^{-F^a \mathbb{X}^a} = G~.
\end{align}
We will take $G_{IJ}=r^2\, \delta_{IJ}$, so that the above constraint is satisfied automatically, because of the antisymmetry of $F_{IJ}^a$. Performing the field redefinition (\ref{FieldRedefinition}) in the $\sigma$-model, we find that the fermionic bilinear part of the deformation (\ref{Deformation}) is precisely absorbed into the kinetic term of the free worldsheet fermions:
\begin{align}
	G_{IJ}\,\psi^I\bar\partial\psi^J~.
\end{align}
Hence, the deformed theory with a curved compact Melvin-like metric:
\begin{align}\label{MelvinMetric}
\begin{split}
	 ds^2 = \left[R^2+r^2(F^2)_{IJ}\hat X^I \hat X^J\right](d\mathbb{X})^2+r^2(d\hat X^I)^2+ 2r^2 & F_{IJ}\hat X^J d\hat X^I d\mathbb{X}~,
\end{split}
\end{align}
with $(F^2)_{IJ} \equiv {F^{K}}_I F_{KJ}$, is reduced to a flat CFT with modified boundary conditions for the (free) worldsheet supercoordinates $\psi^I, X^I$ associated to the toroidal directions. Note that the fibration structure of (\ref{MelvinMetric}) is drastically different than that of the simple example fibrations encountered in Section \ref{SectionMotivation}. They are now  non-trivial compact analogues of Melvin-like fibrations, corresponding to Scherk-Schwarz reduction on $S^1$ from a higher-dimensional flat space, \cite{Melvin:1963qx,Russo:1994cv,Serone:2003sv}.

The shift in the boundary conditions can be seen by observing that, under:
\begin{align}\nonumber
	& \mathbb{X}^a(\sigma^1+2\pi,\sigma^2) = \mathbb{X}^a(\sigma^1,\sigma^2)+2\pi n^a~,\\
	& \mathbb{X}^a(\sigma^1,\sigma^2+2\pi) = \mathbb{X}^a(\sigma^1,\sigma^2)+2\pi \tilde{m}^a~,
\end{align}
the redefined worldsheet fermions $\psi^I$ are rotated as:
\begin{align}\nonumber
	& \psi^I(\sigma^1+2\pi,\sigma^2) = {\left(e^{2\pi  F^a n^a}\right)^I}_J~e^{i\pi(1-a)}\,\psi^J(\sigma^1,\sigma^2)~,\\
	& \psi^I(\sigma^1,\sigma^2+2\pi)= {\left(e^{2\pi  F^a \tilde{m}^a}\right)^I}_J~e^{i\pi(1-b)}\,\psi^J(\sigma^1,\sigma^2)~,
\end{align}
where $n^a,\tilde{m}^a$ are the winding numbers around the $S^1$ directions and $[^a_b]$ are the fermion spin structures of the (unhatted) fermions before the deformation. For our purposes, it will be sufficient to consider only one such circle ($\mathbb{X}^a\equiv\mathbb{X}$), leaving the remaining internal coordinates to parametrize the $T^N\subseteq T^5$. A similar analysis leads to the change in the boundary conditions of the (unhatted) $X^I$-bosons:
\begin{align}\nonumber
	&  X^I(\sigma^1+2\pi,\sigma^2) = {\left(e^{2\pi  F^a n^a}\right)^I}_J\, X^J(\sigma^1,\sigma^2)~,\\
	&  X^I(\sigma^1,\sigma^2+2\pi)= {\left(e^{2\pi  F^a \tilde{m}^a}\right)^I}_J\, X^J(\sigma^1,\sigma^2)~.
\end{align}
Notice that the above field redefinitions and the induced modified boundary conditions act symmetrically in the left- and right- movers. On the other hand, the hatted variables $\hat{\psi},\hat{X}$ continue to have trivial boundary conditions, in agreement with the field redefinitions  (\ref{FieldRedefinitionBoson}),  (\ref{FieldRedefinition}).

The antisymmetric matrix $F^a_{IJ}$ can always be rotated to block-diagonal form by an orthogonal transformation:
\begin{align}
	{(F^a)^I}_J=\left( \begin{array}{r r r r r r}
					0 & f_1^a & 0 & 0 & \ldots & 0 \\
				-f_1^a  & 0	& 0 & 0 & \ldots & 0 \\
				0 & 0 & 0 & f_2^a & \ldots & 0 \\
				0 & 0 & -f_2^a & 0 & \ldots & 0 \\
				\vdots & \vdots & \vdots & \vdots & \ldots & \vdots \\
				\end{array}\right) ~.
\end{align}
We can now define the basis $\Psi^{(f_i)}, Z^{(f_i)}$ that diagonalizes $F^a_{IJ}$, so that each block describes a complex system of free (super-)coordinates. The remaining real (super-)coordinate is associated to the vanishing eigenvalue in the odd-dimensional case. In this (complex) basis, the boundary conditions are expressed as:
\begin{align} \nonumber
	&\Psi^{(f_i)}(\sigma^1+2\pi,\sigma^2) = e^{2\pi i f_i^a n^a} e^{i\pi(1-a)}~ \Psi^{(f_i)}
(\sigma^1,\sigma^2)~,\\
	&\Psi^{(f_i)}(\sigma^1,\sigma^2+2\pi) = e^{2\pi i f_i^a \tilde{m}^a} e^{i\pi(1-a)}~ \Psi^{(f_i)}(\sigma^1,\sigma^2)~,
\end{align}
and similarly for the $Z^{(f_i)}$-bosons. The quantization condition for the $F$-flux now translates into a crystallographic action of the form :
\begin{align}\label{Feigenvalue}
	N f_i \in\mathbb{Z}~.
\end{align}
Thus, the field redefinition (\ref{FieldRedefinition}) maps the curved CFT of real worldsheet fermions $\hat\psi$ with the usual (undeformed) spin structures $[^a_b]$ and with the constant quantized $F$-flux background into a flat CFT of free fermions $\psi$ without $F$-flux, but with modified boundary conditions for each complex fermion system:
\begin{align}
	 \theta[^{a-2f_i n}_{b-2f_i\tilde{m}}] ~.
\end{align}
Of course, the same story is true for the bosonic fields, which can be fermionized as in Section \ref{SectionOrbifolds}. This solvable theory is identified as a freely-acting symmetric $\mathbb{Z}_N$-orbifold. This can be seen immediately by symmetrically extending the orbifold constructions of Section \ref{SectionOrbifolds} to the right-movers and making the identification $f_i^a \leftrightarrow q_f$. This illustrates the equivalence of freely-acting symmetric orbifolds with the conventional `symmetric' Scherk-Schwarz mechanism. 


\subsection{Asymmetric Scherk-Schwarz}
\bigskip

Following the logic of the previous section, we shall now consider an `asymmetric' version of the Scherk-Schwarz mechanism and introduce a deformation rotating only the left-moving internal space. The fermionic part of this deformation will be a left-moving $O(N;\mathbb{Z})$-rotation in the $R$-symmetry lattice:
\begin{align}
	 S^{f}_{\textrm{def}} = -i\int{d^2 z~F_{IJ}^a(\epsilon) {\hat\psi^I \hat\psi^J}\,\tilde{J}^a}~,
\end{align}
where we explicitly include the $\epsilon$-dependence, to treat the case of reflection. Again, $\tilde{J}^a(\bar{z})=i\bar\partial \mathbb{X}^a(\bar{z})$ is a $U(1)_R$-current associated to an $S^1$-coordinate, which is locally factorized from the $T^N$. In our example, we shall identify it with $\tilde{J}^{\mathbb{X}}(\bar{z})=i\bar\partial \mathbb{X}(z)$.

Due to the (local) $\mathcal{N}=1$ worldsheet supersymmetry, this deformation has to be accompanied by a bosonic piece, rotating the left-moving $X^I$-coordinates in $T^N$. As discussed in Section \ref{SectionOrbifolds} on the construction of asymmetric orbifolds, such a left-moving rotation is a lattice automorphism only at special points in moduli space where the left- and right- moving CFTs factorize. We shall therefore take the $T^N$-lattice to lie at the fermionic point and fermionize as before:
\begin{align}
	i\partial \hat{X}^I = i \hat{y}^I \hat\omega^I~.
\end{align}
Rotations acting crystallographically are then naturally identified with the action (\ref{OrbifoldAction}) of the asymmetric $\mathbb{Z}_N$-orbifolds constructed in Section \ref{SectionOrbifolds} and are subject to the same constraints. We are then naturally lead to represent the bosonic $O(N;\mathbb{Z})$-deformation entirely in terms of the free fermions:
\begin{align}
	 S^{b}_{\textrm{def}} &= -i\int{d^2 z~\left( F_{IJ}^a(\epsilon) \hat{y}^I \hat{y}^J+F_{IJ}^a(1)\hat\omega^I\hat\omega^J\right)\,\tilde{J}^a}~.
\end{align}
Hence, repeating the same (fermionic) field redefinitions\footnote{Notice that these field redefinitions do not impinge on the local structure of the non-commutative algebra, since $[\hat{X}^I,\hat{X}^J]=[X^I,X^J]$.} as in the previous section, we can easily identify the $\mathbb{Z}_N$-orbifold action (\ref{OrbifoldAction}) with the rotation induced by the $F$-flux through the field redefinition for a state with unit winding number:
\begin{align}
	P_{IJ}(\epsilon)={\left(e^{2\pi  F^a(\epsilon) }\right)^I}_J~.
\end{align}
In the appropriate basis, this relation fixes the quantization of the $F$-flux (\ref{FQuantiz}) in terms of the allowed $P_{IJ}$-action of consistent asymmetric orbifolds. The full deformation, $ S_{\textrm{def}}=S^f_{\textrm{def}}+ S^b_{\textrm{def}}$, of the $\sigma$-model is a well-defined marginal deformation leading to a solvable CFT of free fermions with modified boundary conditions, equivalent to the freely-acting asymmetric $\mathbb{Z}_N$-orbifolds of Section \ref{SectionOrbifolds}. Indeed, performing the path integral in the $\sigma$-model of free fermions $\{\hat\psi,\hat{y},\hat\omega,\ldots\}$ and adding the deformation $S_{\textrm{def}}$, one recovers the $\mathbb{Z}_N$-orbifold partition function eqs.~(\ref{GeneralPartition})-(\ref{Partition3}) of Section \ref{SectionOrbifolds}.

The $U(1)_R$-flux $F^a_{IJ}(\epsilon)$ defined above is associated to an $O(N;\mathbb{Z})$-rotation which is not part of the group of large diffeomorphisms of the $T^N$-torus and is, hence, identified as a non-geometric flux. Notice that $F_{IJ}^a$ is only antisymmetric in the two toroidal indices $I,J$. We see from our construction that, even though the flux $F$ lacks a geometric interpretation, it has a clear description in terms of the underlying CFT. The equivalence between the asymmetric orbifold picture and the `asymmetric' analogue of the Scherk-Schwarz mechanism allows one to gain considerable insight into the CFT description of non-geometric $Q$-fluxes. The presence of such non-geometric fluxes is precisely the source of the emergent non-commutativity (\ref{GeneralAlgebra}) in the toroidal coordinates.


~\\
\subsection*{Acknowledgements}

We are grateful to I. Bakas, E. Dudas,  C. Lazaroiu, K. Siampos, N. Toumbas, M. Tsulaia and especially to C. Kounnas for interesting comments and discussions. C.C. is thankful to Ecole Polytechnique (CPhT) and I.F. is thankful to Ecole Normale Sup\'erieure (LPTENS) for hospitality during the initial stages of this project. D.L. would also like to thank the Simons Center for Geometry and Physics for its hospitality. The authors are grateful to the CERN Theory Unit for hospitality during the final stages of this work. The research of C.C. is supported by a grant of the Romanian National Authority for Scientific Research, CNCS - UEFISCDI, project number PN-II-RU-PD-2011-3-0033. This work is also supported by the Deutsche Forschungsgemeinschaft, by the DFG Transregional Collaborative Research Centre TRR 33 and the DFG cluster of excellence ``Origin and Structure of the Universe".


\bigskip
\medskip

\bibliographystyle{unsrt}

\vfill\eject
\end{document}